\documentclass{article}

\usepackage{graphicx}
\usepackage{amsmath}

\title{Epidemiological geographic profiling for a meta-population network}

\author{Yoshiharu Maeno \\ Social Design Group \\ email: maeno.yoshiharu@socialdesigngroup.com}

\begin{document}

\maketitle

\begin{abstract}
Epidemiological geographic profiling is a statistical method for making inferences about likely areas of a source from the geographical distribution of patients. Epidemiological geographic profiling algorithms are developed to locate a source from the dataset on the number of new cases for a meta-population network model. It is found from the WHO dataset on the SARS outbreak that Hong Kong remains the most likely source throughout the period of observation. This reasoning is pertinent under the restricted circumstance that the number of reported probable cases in China was missing, unreliable, and incomprehensive. It may also imply that globally connected Hong Kong was more influential as a spreader than China. Singapore, Taiwan, Canada, and the United States follow Hong Kong in the likeliness ranking list.
\end{abstract}

\section{Introduction}

The spreading of an infectious disease is a complex stochastic reaction-diffusion process in a spatially heterogeneous medium. Planning efficiently targeted public health intervention is a pressing task to public health authorities when the authorities detect an outbreak of an infectious disease. The task is urgent particularly if the authorities suspect that the disease is such a new airborne disease as SARS in 2003, which spreads without delay across regional borders by an aviation transportation network. For this purpose, it is essential to estimate and predict the geographical distribution of patients. A source is a significant initial condition in reproducing the time evolution of the distribution in a multiregional epidemic. The source is the location of an index case patient.

A question arises here. Can the source of a multiregional epidemic be located retrospectively when a possible outbreak is detected after a long time?

In this study, epidemiological geographic profiling algorithms are developed for a standard epidemiological SIR compartment model in a meta-population network. Epidemiological geographic profiling is a statistical method for making inferences about likely areas of a source from the geographical distribution of patients. The model is a special case of a stochastic reaction-diffusion process. The entire population is sub-divided into distinct sub-populations in multiple geographical regions. The state of an individual person in the sub-population changes from the susceptible state to the infectious and removed (recovered) states. The performance of the algorithms is investigated with computationally synthesized datasets. The datasets are a snapshot geographical distribution of new cases in a time interval. The algorithms are applied to the WHO dataset on the SARS outbreak. Some findings on the dataset is presented.

\section{Related works}

Geographic profiling is a statistical method in criminology. It aids crime investigators in making inference about likely areas of such a home base as the residence of a serial murderer, rapist, or arsonist from crime sites in a two dimensional topographic space. The possibility of a crime decays as a crime site becomes more distant from the home base because the offender does not want to consume much effort, time, and money for a long travel. The distance is either the displacement between the crime site and the home base or the length along the travel route. Criminologists proposed geographic profiling algorithms with distinctive decay functions to associate the distance from the home base with the possibility of a crime\cite{Can00}, \cite{Ros93}. The algorithm generates a likeliness map which indicates the possibility that the area includes the offender's home base.

Recently geographic profiling is applied to the investigation of an infectious disease outbreak. It aids public health investigators in making inference about likely areas of such a source as the residence of an index case patient from the geographical distribution of patients. A similar likeliness map is generated either by the geographic profiling algorithm in criminology\cite{Toh13}, \cite{LeC12}, \cite{LeC11}, or by an algorithm developed for a mathematical model for
the spread of an infectious disease\cite{Bus13}.

Such mathematical models as dilation, diffusion, and contact processes describe human-to-human transmission of an infectious disease in a social network. Geographic profiling for a social network is called a patient-zero problem. Large unbiased betweenness is an excellent indicator of a source in a social network\cite{CesC11}. A dynamic message-passing algorithm is developed to compute a source probability density function in mean-field approximation\cite{Lok14}. An analytic method is invented to compute an approximate source probability density function for a broad class of network topologies\cite{Ant15}. A community partition method in a reverse propagation model is proposed to locate multiple sources\cite{Zan14}. A variant of a belief propagation algorithm works effectively when observations are incomplete (partially missing) and erroneous with noise\cite{Alt14}.

But none of those studies address a meta-population network model.

\section{Problem}

The entire population is sub-divided into distinct sub-populations in multiple geographical regions. The number of geographic regions is $N$. The geographical regions are nodes $n_{i} \ (i=0,1,\dots,N-1)$. The transportation between geographical regions is a link between nodes. The topology of the transportation is represented by an $N \times N$ adjacency matrix $\mbox{\boldmath{$l$}}$. A link from $n_{i}$ to $n_{j}$ is present if the $i$-th row and $j$-th column element $l_{ij}$ is $1$, and absent if $l_{ij}$ is $0$. The index case patients appear at the source node $n_{{\rm src}} \in \{ n_{i} \}$ at $t=0$. When an outbreak is detected later, observations are made at the all nodes at $t=t_{{\rm obs}} > 0$. The dataset is a vector quantity $\mbox{\boldmath{$D$}}$ whose $i$-th element is the number of new cases at $n_{i}$ in a time interval $\Delta t$. $D_{i}=\Delta J_{i}(t_{{\rm obs}})=J_{i}(t_{{\rm obs}}+\Delta t)-J_{i}(t_{{\rm obs}})$. $J_{i}(t)$ is the cumulative number of new cases at $n_{i}$ until $t$. An example of the dataset is a bundle of the daily reports ($\Delta t = 1$ day) on new cases from hospitals in nearby cities.

The problem is to locate $n_{{\rm src}}$ among $\{ n_{i} \}$ only from given adjacency matrix $\mbox{\boldmath{$l$}}$ and dataset $\mbox{\boldmath{$D$}}$. Neither $\mbox{\boldmath{$J$}}(t_{{\rm obs}})$ nor $t_{{\rm obs}}$ are known because the time when the index case patients appear is not known. Nothing is known about transmission parameters, sub-populations, and relevant initial conditions.

\section{method}
\label{method}

The likeliness score as a source node is calculated for the all nodes. The likeliness ranking of a node is the descending order of the likeliness score of the node. The node at the top in the likeliness ranking list is the most probable source node. The likeliness score $s_{i}$ of $n_{i}$ is defined by a scalar product of two vectors in eq.(\ref{score}). The quantity$\mbox{\boldmath{$w$}}_{i}$ is a decay vector of the distance between nodes.
\begin{equation}
s_{i} = \frac{\mbox{\boldmath{$w$}}_{i}}{|\mbox{\boldmath{$w$}}_{i}|} \cdot \frac{ \Delta \mbox{\boldmath{$J$}} (t_{{\rm obs}}) }{|\Delta \mbox{\boldmath{$J$}} (t_{{\rm obs}})|}.
\label{score}
\end{equation}

The distance is an integer hop count, which is the smallest number of intermediate links along a route between the nodes. The decay function is given by eq.(\ref{decay}). The distance between $n_{i}$ and $n_{j}$ is $d_{ij}$.
\begin{equation}
\mbox{\boldmath{$w$}}_{i} = (w(d_{i0}), w(d_{i1}), \dots, w(d_{i\ N-1}).
\label{decay}
\end{equation}

The algorithm is characterized by the functional forms of the decay function in eq.(\ref{naive}) through (\ref{exponential}). Four functional forms are investigated in this study. The first function is a naive function in eq.(\ref{naive}) where $\delta$ is a Kronecker's symbol. The number of infectious persons at neighboring nodes ($d \geq 1$) is not significant. If the number of infectious persons is larger, it is more probable that the node is a source.
\begin{equation}
w(d) = \delta_{d0}.
\label{naive}
\end{equation}

The second function is a power function in eq.(\ref{power}). A network is a collection of chains approximately if the network is non-disjoint but very sparse. The mean number of a time-evolving variable in a similar Langevin equation for a chain network includes a power function of the distance from the source node to the node\cite{Mae13}. This dependence is represented concisely by eq.(\ref{power}).
\begin{equation}
w(d) = \frac{\pi^{d}}{d!}.
\label{power}
\end{equation}

The third function is a polynomial function in eq.(\ref{polynomial}). The delay is relatively slow. As $\rho \rightarrow \infty$, the function converges to the naive function in eq.(\ref{naive}). A similar function is applied to the Rossmo algorithm for two dimensional criminological geographic profiling\cite{Ros93}.
\begin{equation}
w(d) = \frac{1}{(d+1)^{\rho}}.
\label{polynomial}
\end{equation}

The fourth function is an exponential function in eq.(\ref{exponential}). The decay is relatively fast. As $\sigma \rightarrow \infty$, the function converges to the naive function in eq.(\ref{naive}). A similar function is employed by the Canter algorithm for two dimensional criminological geographic profiling\cite{Can00}.
\begin{equation}
w(d) = \frac{1}{e^{\sigma d}}.
\label{exponential}
\end{equation}

\section{Result}

\subsection{Synthesized dataset}

In a standard epidemiological SIR compartment model, the state of a person changes from a susceptible state, through an infectious state, to a removed (recovered) state. The time dependent variables $S_{i}(t)$, $I_{i}(t)$, and $R_{i}(t)$ are the number of susceptible persons, infectious persons, and removed persons at $n_{i}$ at $t$. Their time evolution is described by a system of Langevin equations\cite{Huf04}, \cite{Mae11}. The time evolution of $I_{i}(t)$ is given by eq.(\ref{dI/dt}). The fluctuation terms $\xi(t)$ represent Gaussian white noise.
\begin{eqnarray}
\frac{{\rm d} I_{i}(t)}{{\rm d} t} &=& \frac{\alpha S_{i}(t) I_{i}(t)}{S_{i}(t)+I_{i}(t)+R_{i}(t)} - \beta I_{i}(t) 
+ \sum_{j \neq i} \gamma_{ji} I_{j}(t) - \sum_{j \neq i} \gamma_{ij} I_{i}(t) \nonumber \\
&+& \sqrt{\frac{\alpha S_{i}(t) I_{i}(t)}{S_{i}(t)+I_{i}(t)+R_{i}(t)}} \xi^{[\alpha]}_{i}(t) 
- \sqrt{\beta I_{i}(t)} \xi^{[\beta]}_{i}(t) \nonumber \\
&+& \sum_{j} \sqrt{\gamma_{ji} I_{j}(t)} \xi^{[\gamma]}_{ji}(t) - \sum_{j} \sqrt{\gamma_{ij} I_{i}(t)} \xi^{[\gamma]}_{ij}(t).
\label{dI/dt}
\end{eqnarray}

The parameter $\alpha$ is the probability of an infectious person contacting a person and infecting the person per a unit time, and $\beta$ is the probability of an infectious person being removed per a unit time. The reproductive ratio\cite{Ril03} is defined by $r=\alpha/\beta$. The parameter $\gamma_{ij}$ is the probability of a person moving from $n_{i}$ to $n_{j}$ per a unit time. It is known empirically that $\gamma_{ij}$ is determined from the adjacency matrix by an empirical law in eq.(\ref{gammacalc}) where $k_{i}$ is the nodal degree of $n_{i}$. This law is known valid for an aviation transportation network\cite{Bar04}. 
\begin{eqnarray}
\gamma_{ij}= \frac{l_{ij} \sqrt{k_{i} k_{j}}}{\sum_{j=0}^{N-1} l_{ij} \sqrt{k_{i} k_{j}}} \gamma.
\label{gammacalc}
\end{eqnarray}

The time evolution of $J_{i}(t)$ is given by eq.(\ref{dJ/dt}).
\begin{eqnarray}
\frac{{\rm d} J_{i}(t)}{{\rm d} t} = \alpha I_{i}(t) + \sqrt{\alpha I_{i}(t)} \xi^{[\alpha]}_{i}(t).
\label{dJ/dt}
\end{eqnarray}

\begin{figure}
\includegraphics[scale=0.35,angle=-90]{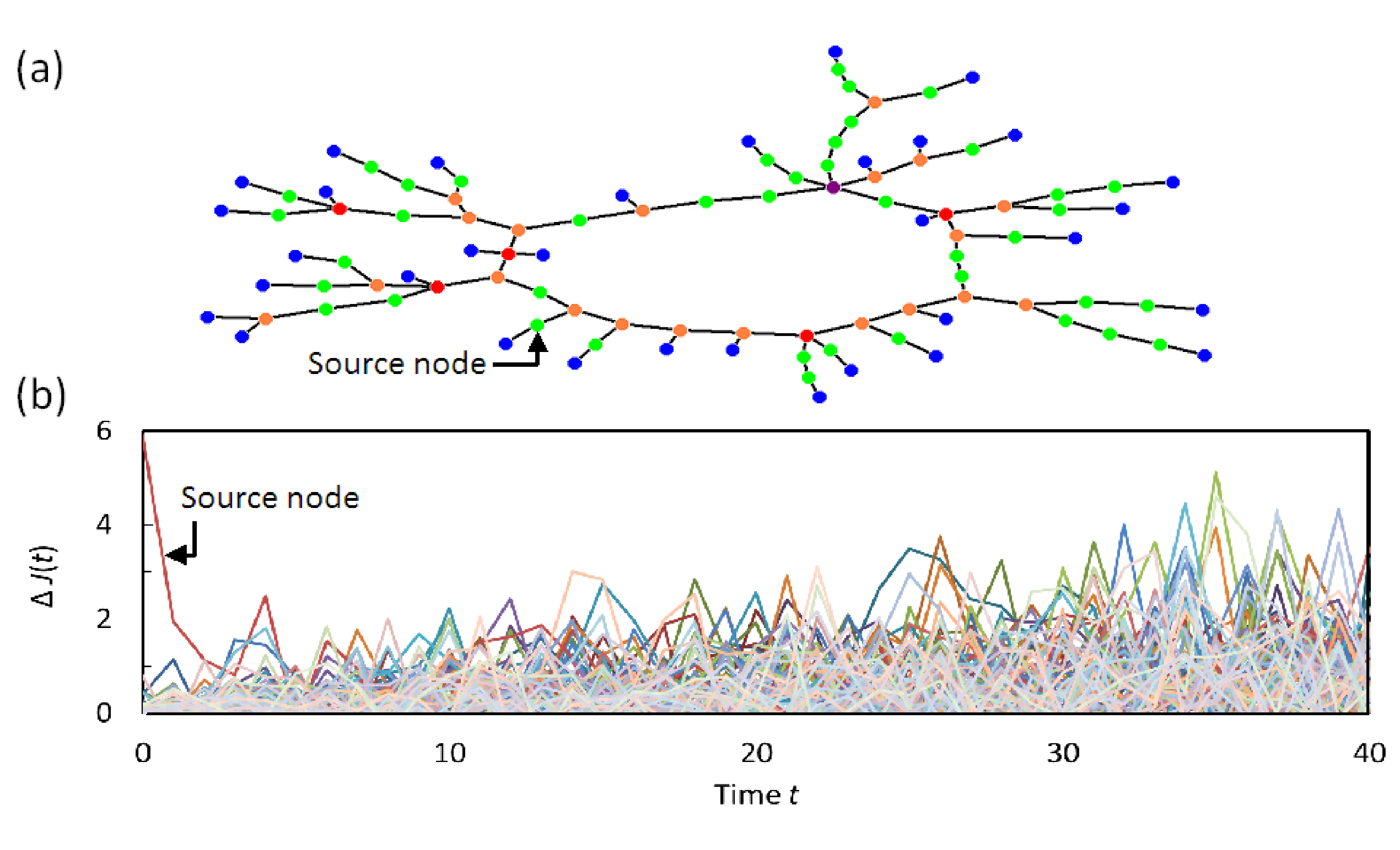}
\caption{(a) Example of a transportation network with $N=100$ when $\bar{k}=2$. (b) Example dataset $\mbox{\boldmath{$D$}} = \Delta \mbox{\boldmath{$J$}}(t)$ as a function of $t$ when $\alpha=0.16$, $\beta=0.04$ ($r=4$), and $\gamma=0.2$ for the network shown in (a).}
\label{file1s}
\end{figure}

The algorithms in \ref{method} are tested with synthesized datasets. Datasets are synthesized by solving eq.(\ref{dI/dt}) through eq.(\ref{dJ/dt}) numerically with a pseudo random number generator and assembling $\Delta J_{i}$ at $t_{{\rm obs}}$ under given experimental conditions. The topology of a transportation network is an Erdos-Renyi random network with $N=100$ nodes. Figure \ref{file1s} (a) shows an example when the average nodal degree is $\bar{k}=\sum_{i=0}^{N-1} k_{i}/N=2$. Figure \ref{file1s} (b) shows an example dataset as a function of $t$ when $\alpha=0.16$, $\beta=0.04$ ($r=4$), and $\gamma=0.2$. The source node is chosen randomly. The number of index case patients at the source node is $I_{{\rm src}}(0)=20$. There are not any patients at the rest of the nodes. The entire population is $10^{8}$.

A hit score\cite{LeC11} quantifies the performance of an algorithm. It is the cumulative search area as a fraction of the total area, which is investigated, according to the likeliness ranking of nodes, to find the source node. The area is the number of nodes for a meta-population network model. The hit score ranges from 0 to 1. A smaller value of the hit score means a better performance of the algorithm. The hit score is 0.5 for the random search. The hit score $H$ is defined by eq.(\ref{hit score}). The likeliness score of the source node is $s_{{\rm src}}$.
\begin{equation}
H = \frac{1}{N} |\{ i|s_{i} \geq s_{{\rm src}} \}|.
\label{hit score}
\end{equation}

\begin{figure}
\includegraphics[scale=0.42,angle=-90]{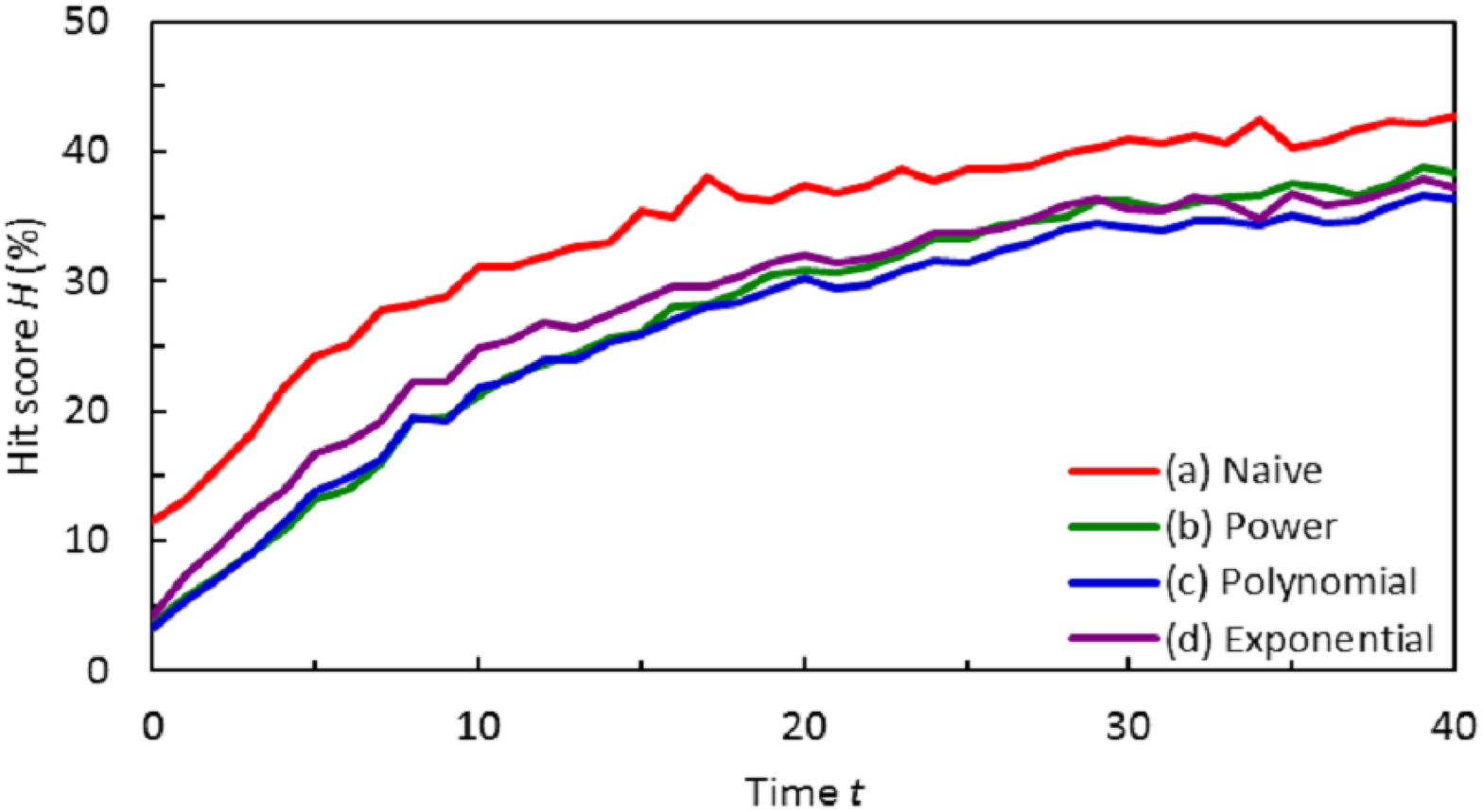}
\caption{Hit score $H$ for the four algorithms ((a) naive, (b) power, (c) polynomial, (d) exponential) as a function of time $t$ when $\bar{k}=2$, $\alpha=0.11$, $\beta=0.09$ ($r=1.2$), and $\gamma=0.2$. }
\label{file2s}
\end{figure}

The values for $\pi$ in eq.(\ref{power}), $\rho$ in eq.(\ref{polynomial}), and $\sigma$ in eq.(\ref{exponential}) are chosen so that $H$ can be the smallest and the performance of individual algorithms can be the most excellent. The optimal values are $\hat{\pi}=2$, $\hat{\rho}=0.5$, and $\hat{\sigma}=0.05$. Figure \ref{file2s} shows $H$ for the four algorithms as a function of $t$ when $\bar{k}=2$ and $\alpha=0.11$, $\beta=0.09$ ($r=1.2$), and $\gamma=0.2$. The hit scores in the graph are the average values over 1000 different topologies of a transportation network. The algorithm with a polynomial function is the most excellent of the four algorithms. The hit score increases and the all algorithms come closer to the random search as time elapses.

\begin{figure}
\includegraphics[scale=0.42,angle=-90]{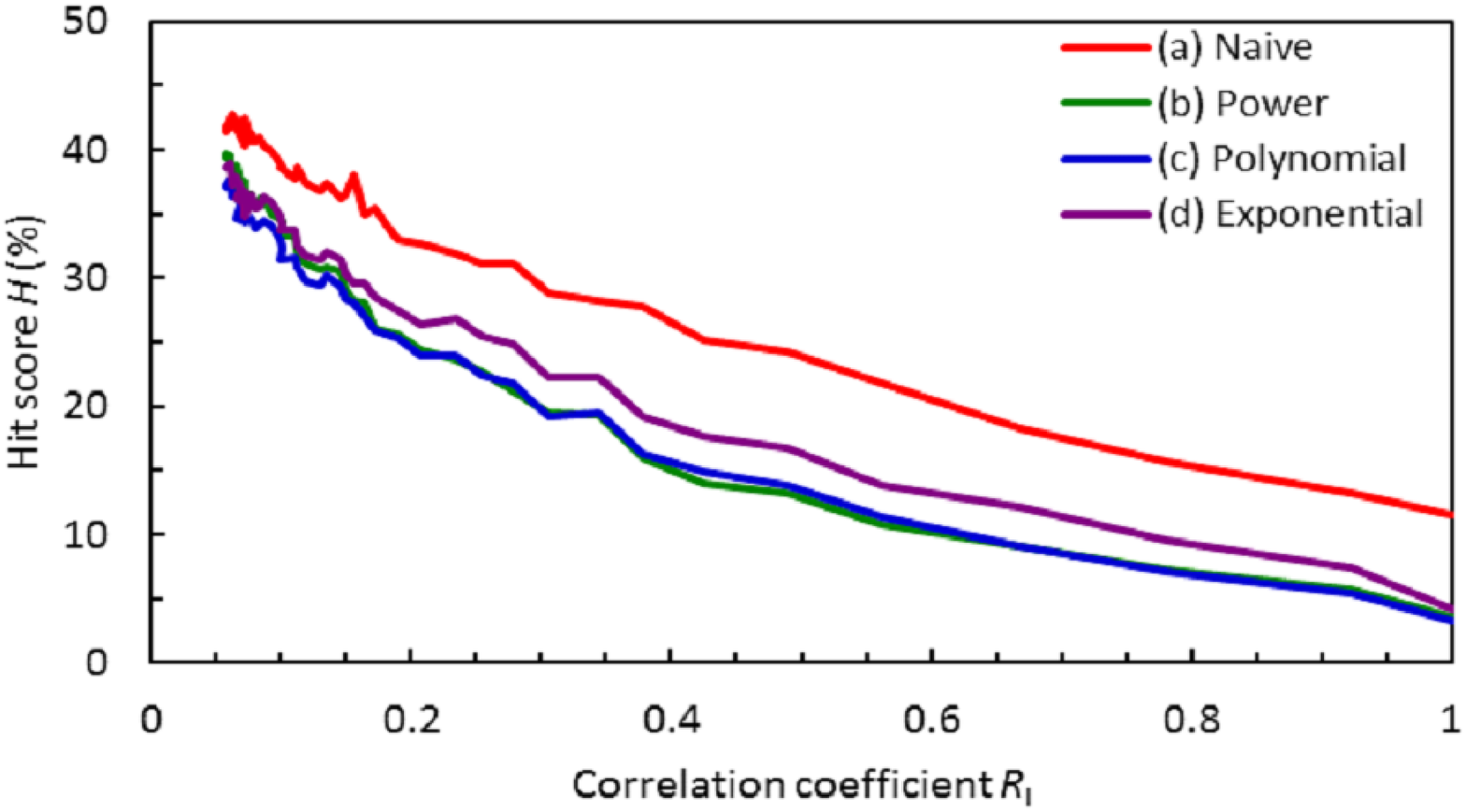}
\caption{Hit score $H$ for the four algorithms ((a) naive, (b) power, (c) polynomial, (d) exponential) as a function of the correlation coefficient $R_{I}$ when $\bar{k}=2$, $\alpha=0.11$, $\beta=0.09$ ($r=1.2$), and $\gamma=0.2$.}
\label{file3s}
\end{figure}

Figure \ref{file3s} shows $H$ for the four algorithms as a function of the correlation coefficient $R_{I}$ under the same experimental conditions as those of Figure \ref{file2s}. $R_{I}$ is the Pearson's product-moment correlation coefficient between the present geographical distribution of patients $\mbox{\boldmath{$I$}}(t)$ at $t$ and the initial distribution $\mbox{\boldmath{$I$}}(0)$. The hit score is a monotonically decreasing function of the correlation coefficient. The stochastic process destroys the trace of the initial conditions in the present state of sub-populations. The trace cannot be restored any more after a long time under this experimental condition.

\begin{figure}
\includegraphics[scale=0.42,angle=-90]{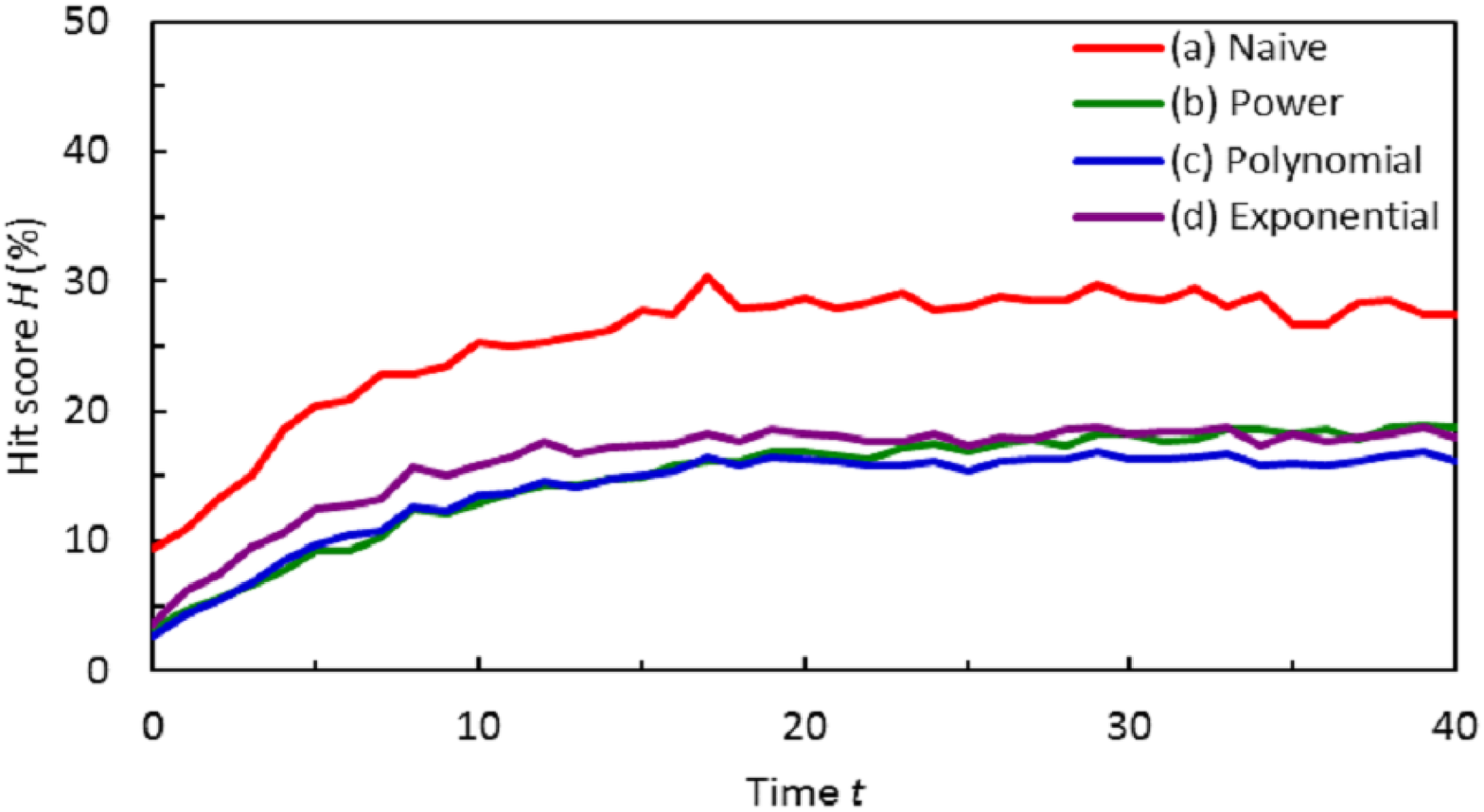}
\caption{Hit score $H$ for the four algorithms ((a) naive, (b) power, (c) polynomial, (d) exponential) as a function of time $t$ when $\bar{k}=2$, $\alpha=0.133$, $\beta=0.067$ ($r=2$), and $\gamma=0.2$. }
\label{file4s}
\end{figure}

Figure \ref{file4s} shows $H$ for the four algorithms as a function of $t$ when $\bar{k}=2$, $\alpha=0.133$, $\beta=0.067$ ($r=2$), and $\gamma=0.2$. The polynomial function is still the most excellent. The performance converges to $H \approx 0.15$ as time elapses. The algorithms work excellently. The trace of the initial conditions is not destroyed under this experimental condition because large reproductive ratio makes the number of infectious persons keep on growing at the source node.

\begin{figure}
\includegraphics[scale=0.42,angle=-90]{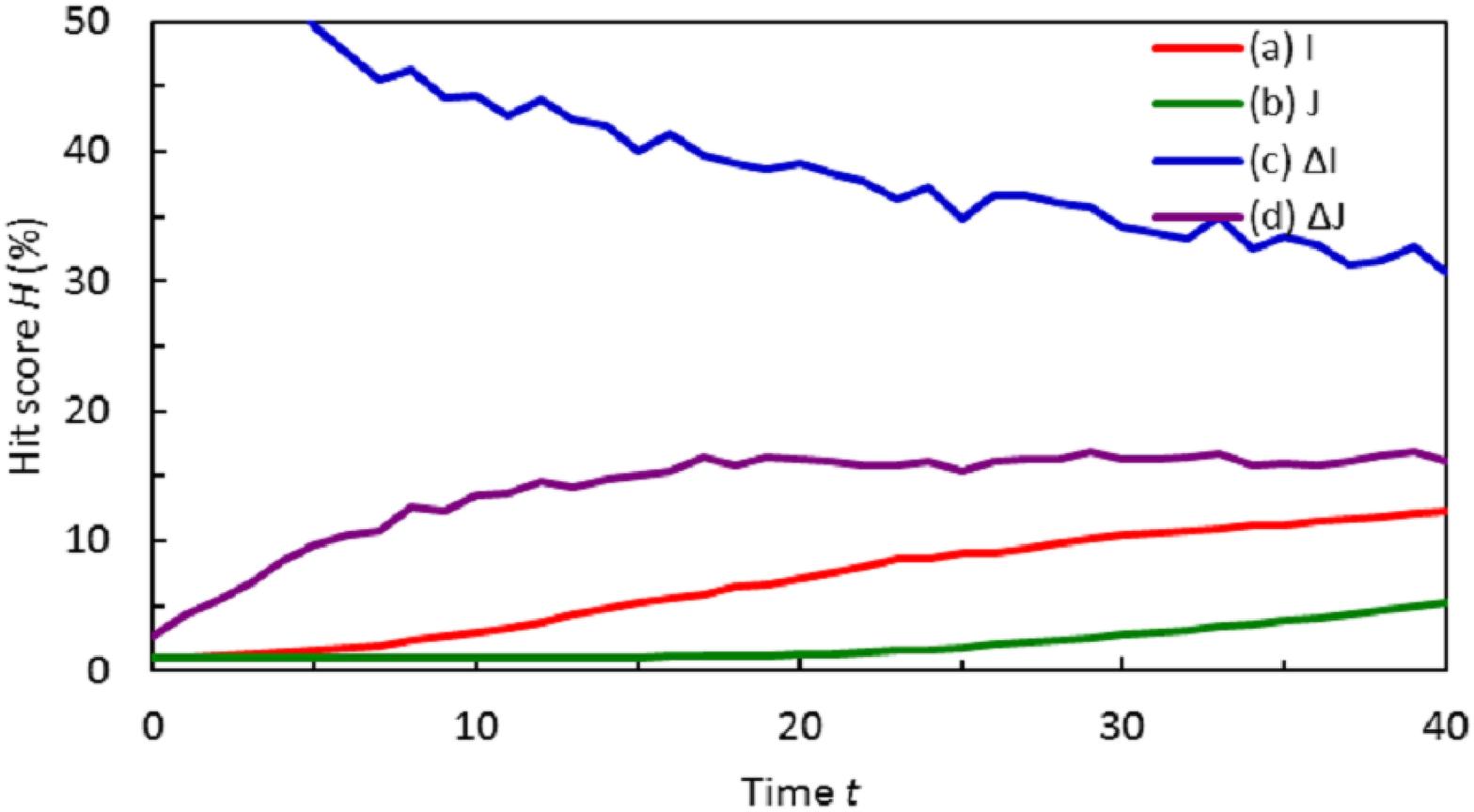}
\caption{Hit score $H$ for the polynomial algorithm as a function of time $t$ when $\bar{k}=2$, $\alpha=0.133$, $\beta=0.067$ ($r=2$), $\gamma=0.2$, the dataset is (a) $\mbox{\boldmath{$I$}}$, (b) $\mbox{\boldmath{$J$}}$, (c) $\Delta \mbox{\boldmath{$I$}}$, (d) $\Delta \mbox{\boldmath{$J$}}$. }
\label{file5s}
\end{figure}

Figure \ref{file5s} shows $H$ for the polynomial function under the same experimental conditions as those of Figure \ref{file4s} when the dataset is $\mbox{\boldmath{$I$}}$, $\mbox{\boldmath{$J$}}$, $\Delta \mbox{\boldmath{$I$}}$, and $\Delta \mbox{\boldmath{$J$}}$. $\mbox{\boldmath{$J$}}$ is the most informative in locating the source while $\Delta \mbox{\boldmath{$I$}}$ is not informative at all. If either $\mbox{\boldmath{$J$}}$ or $\mbox{\boldmath{$I$}}$ were observable directly, the hit score would be as small as a third of that for $\Delta \mbox{\boldmath{$J$}}$. 

\begin{figure}
\includegraphics[scale=0.42,angle=-90]{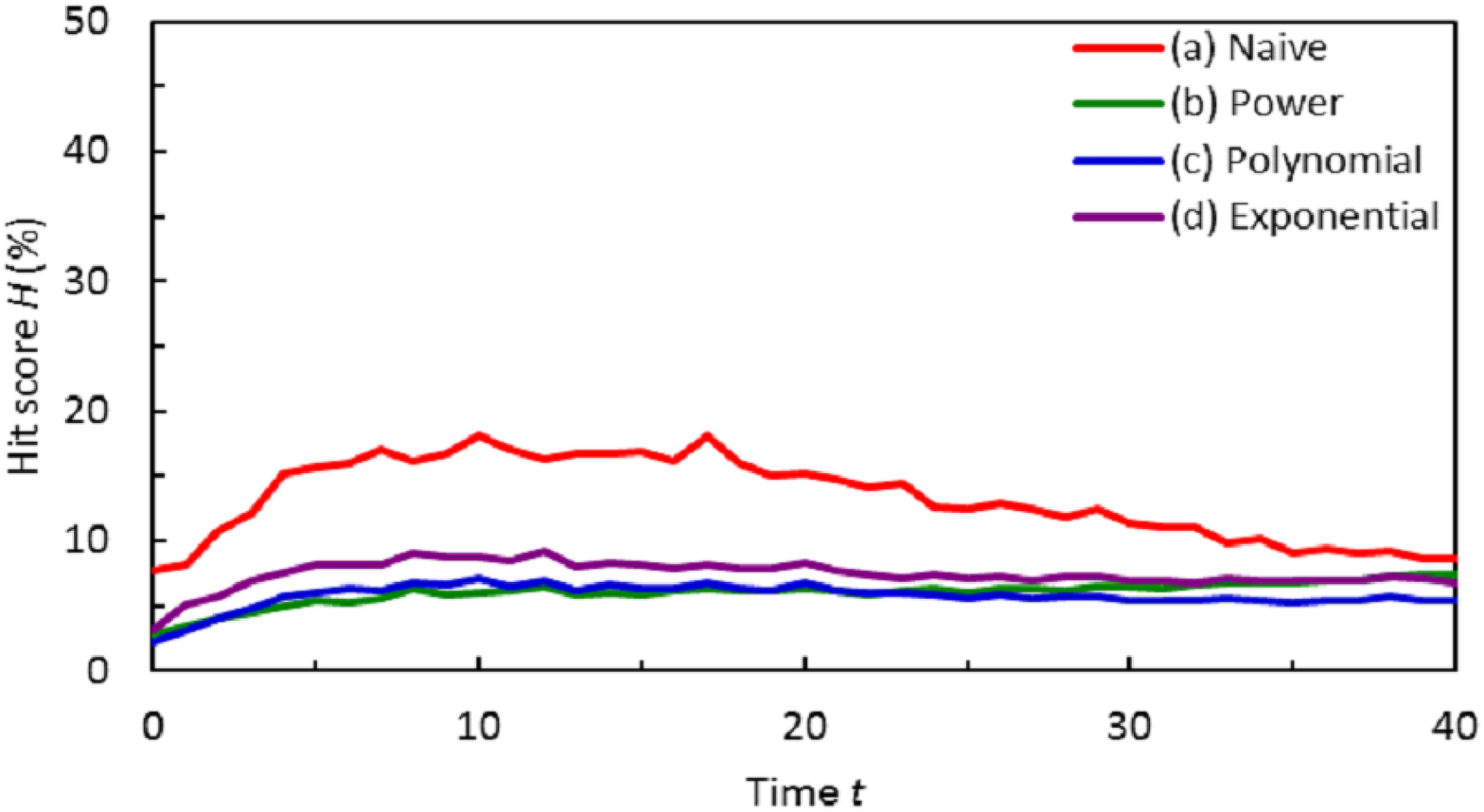}
\caption{Hit score $H$ for the four algorithms ((a) naive, (b) power, (c) polynomial, (d) exponential) as a function of time $t$ when $\bar{k}=2$, $\alpha=0.16$, $\beta=0.04$ ($r=4$), and $\gamma=0.2$.}
\label{file6s}
\end{figure}

Figure \ref{file6s} shows $H$ for the four algorithms as a function of $t$ when $\bar{k}=2$, $\alpha=0.16$, $\beta=0.04$ ($r=4$), and $\gamma=0.2$. The algorithm with a polynomial function is still the most excellent. The hit score does not increase at all as time elapses. It remains at a small value $H \approx 0.05$.

\begin{figure}
\includegraphics[scale=0.42,angle=-90]{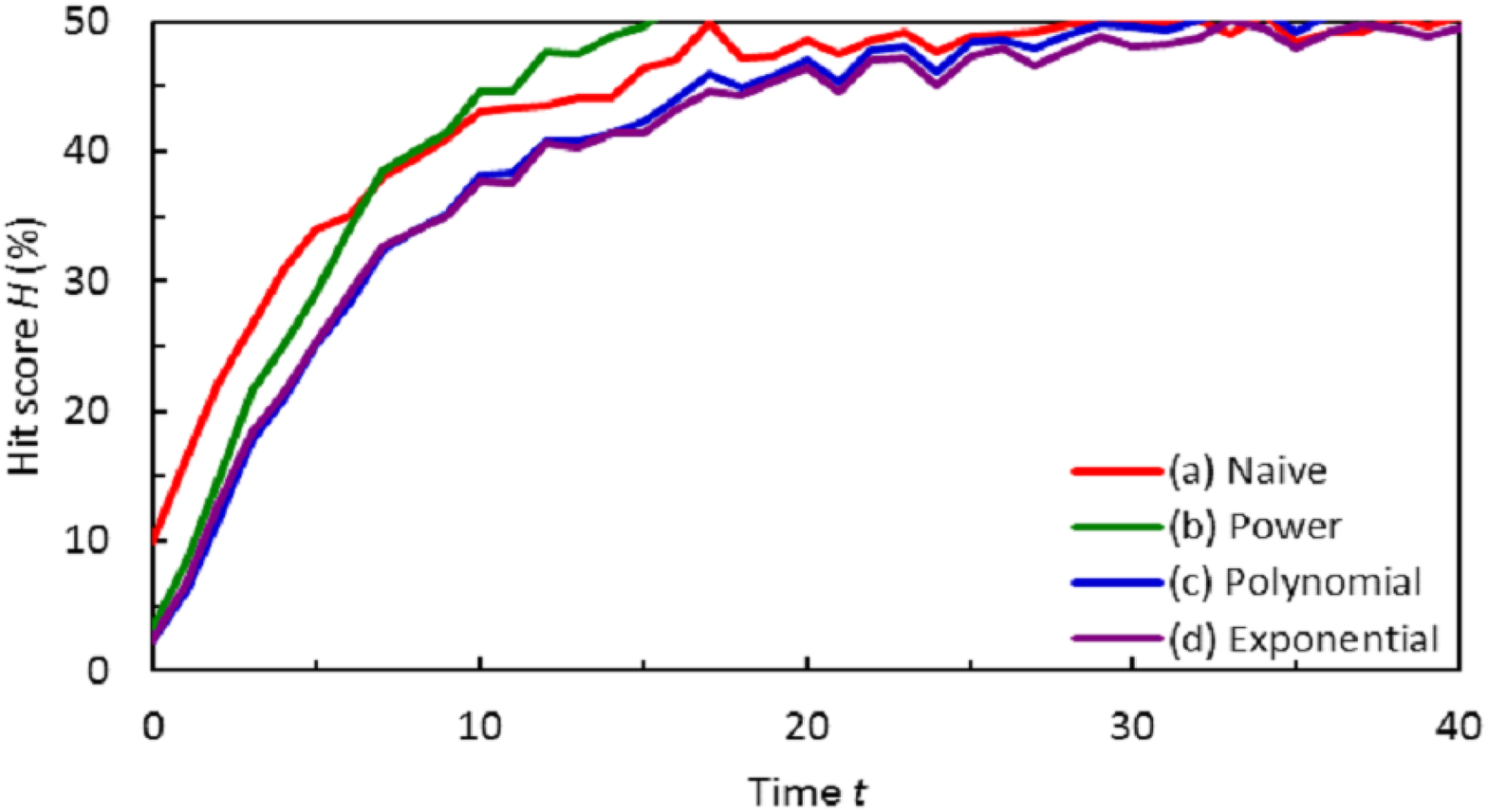}
\caption{Hit score $H$ for the four algorithms ((a) naive, (b) power, (c) polynomial, (d) exponential) as a function of time $t$ when $\bar{k}=4$, $\alpha=0.16$, $\beta=0.04$ ($r=4$), and $\gamma=0.2$.}
\label{file7s}
\end{figure}

Figure \ref{file7s} shows $H$ for the four algorithms as a function of $t$ when $\bar{k}=4$, $\alpha=0.16$, $\beta=0.04$ ($r=4$), and $\gamma=0.2$. The hit score increases more quickly as time elapses. The infectious disease spreads to multiple geographical regions quickly because of the large nodal degree. Quick relaxation to the equilibrium state ensues. Along with the stochastic process, this effect destroys the trace of the initial conditions.

\subsection{Real dataset}

SARS is a respiratory disease in humans caused by the SARS corona-virus. The epidemic of SARS appears to have started in Guangdong of south China in November 2002. SARS spread from the Guangdong to Hong Kong in early 2003, and eventually nearly 40 countries around the world by July\cite{Lip03}, \cite{Ril03}. The WHO archives the cumulative number of reported probable cases\footnote{World Health Organization, Cumulative number of reported probable cases of SARS, http://www.who.int/csr/sars/country/en/index.html (2003).}. The archive is transformed to the dataset $\mbox{\boldmath{$D$}} = \Delta \mbox{\boldmath{$J$}}$. The cumulative number of reported cases is not the cumulative number of new cases until $t$. Just the increase in the cumulative number of reported cases is significant to this study. The target geographical regions are those where five or more cases had been reported in a month since March 17. They are Canada (CAN), France (FRA), the United Kingdom (GBR), Germany (GER), Hong Kong (HKG), Malaysia (MAS), Taiwan (ROC), Singapore (SIN), Thailand (THI), the United States (USA), and Vietnam (VIE). China is not included in the target geographical regions because the the number of reported probable cases there was missing, unreliable, and incomprehensive at the time of observation. The number of geographical regions is $N=11$. The time interval between observations is $\Delta t=1$ day.

\begin{figure}
\includegraphics[scale=0.42,angle=-90]{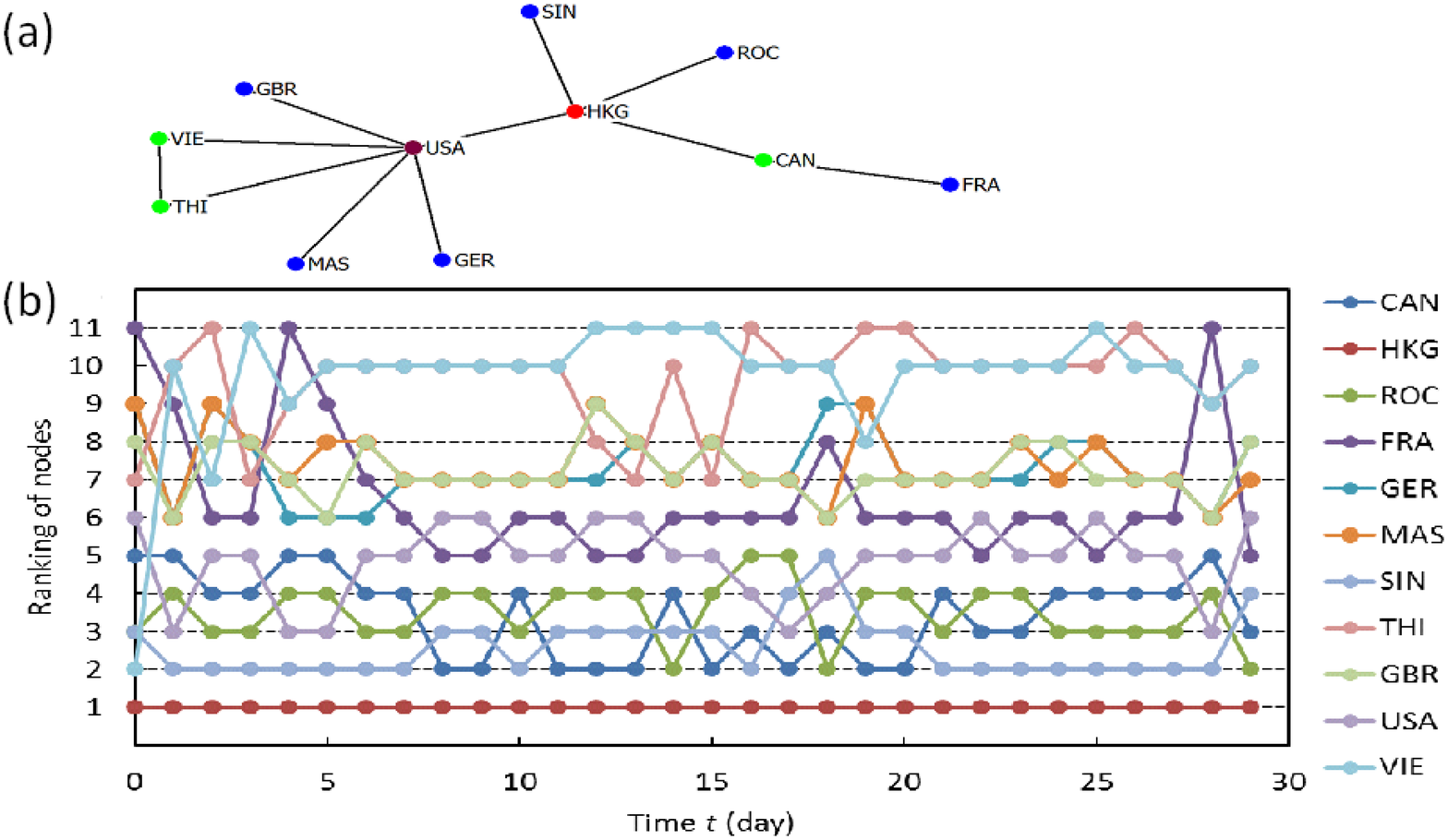}
\caption{(a) Topology of the avian transportation network among Canada (CAN), France (FRA), the United Kingdom (GBR), Germany (GER), Hong Kong (HKG), Malaysia (MAS), Taiwan (ROC), Singapore (SIN), Thailand (THI), the United States (USA), and Vietnam (VIE). (b) Likeliness ranking of the all nodes as a function of $t$.}
\label{file8s}
\end{figure}

The effectively decisive topology of an avian transportation network is discovered from the dataset \cite{Mae10}. This topology is represented by an adjacency matrix $\mbox{\boldmath{$l$}}$ in Figure \ref{file8s} (a). Thiland and Veitnam share a topologically equivalent position. Malaysia, the United Kingdom, and Germany are also equivalent. Those may have the same likeliness ranking.

Figure \ref{file8s} (b) shows the likeliness ranking of the all nodes as a function of $t$. The decay function in eq.(\ref{score}) is a polynomial function in eq.(\ref{polynomial}) with $\hat{\rho}=0.5$. The first observation (on March 17) is made at $t=0$, that is, $t_{{\rm obs}}=0, 1, 2, \cdots$. The index case patient appears and the consequent spread starts at an unknown earlier time. Hong Kong remains the most probable source node throughout the month. Singapore, Taiwan, and Canada follow Hong Kong in the likeliness ranking list. The United States, which is the largest hub node, is not so likely to be a source node as those four nodes. Interestingly, the dataset implies that such Southeast Asian regions as Malaysia, Thiland, Vietnam are less likely to be a source node than such European regions as France, the United Kingdom, Germany. Those results are consistent with the fact from an official report\footnote{SARS Expert Committee (Hong Kong), SARS in Hong Kong: from experience to action, http://www.sars-expertcom.gov.hk/english/reports/reports/reports\_fullrpt.html (2003).} that SARS spreads from Hong Kong to Singapore and Taiwan. 

\section{Discussion}
The epidemiological geographic profiling algorithm with a polynomial decay function works most excellently for computationally synthesized datasets. The hit score remains at a small value as far as the average nodal degree is small or the reproductive ratio is large.   It is found from the WHO dataset on the SARS outbreak that Hong Kong remains the most likely source throughout the period of observation. This reasoning is pertinent under the restricted circumstance that the number of reported probable cases in China was missing, unreliable, and incomprehensive. It may also imply that globally connected Hong Kong was more influential as a spreader than China. Singapore, Taiwan, Canada, and the United States follow Hong Kong in the likeliness ranking list.

The number of infectious persons $\mbox{\boldmath{$I$}}(s)$ at time $s$ governs the time evolution of the epidemic for $t \geq s$. Integrating a system of Langevin equations in eq.(\ref{dI/dt}) is a forward problem. The forward problem could be solved in a straightforward manner if the initial conditions and boundary conditions were known. Acquiring these conditions from observation is an inverse problem. The inverse problem ranges from finding the initial values of time-evolving variables, estimating the transmission parameters, and discovering the topology of transportation network to detecting change points in the parameters or the topology. Once the initial conditions are acquired by the epidemiological geographic profiling in this study, public health authorities can reproduce the time evolution of a multiregional epidemic. The investigators can quantify the risk of infection at any future time in every geographical region. They may find that a heavily populated urban center is on the verge of a pandemic. Protecting the center from infection by immediate intensive border screening may be reasonable public health intervention. They may also find that a city is a potential super-spreader. Isolating the infectious city from the neighboring susceptible cities may be first-aid public health intervention.

The epidemiological geographic profiling algorithm is a computationally efficient heuristic. A system of Langevin equations is equivalent approximately to a Kolmogorov forward equation in probability theory, or a Fokker-Planck equation in statistical physics, for a multivariate probability density function\cite{Mae10}. Another problem formulation is solving a Kolmogorov backward equation subject to the observation as the final conditions. This approach is more exact than the heuristic, but often less tractable. Resolving such complexity is for future studies.


\begin{thebibliography}{[MT1]}

\bibitem[Antulov-Fantulin 2015]{Ant15} N. Antulov-Fantulin, A. Lan\u{c}i\'{c}, T. \u{S}muc, H. \u{S}tefan\u{c}i\'{c}, M. \u{S}iki\'{c}, Identification of patient zero in static and temporal networks: Robustness and limitations, Physical Review Letters {\bf 114}, 248701 (2015).

\bibitem[Altarelli 2014]{Alt14} F. Altarelli, A. Braunstein, L. Dall'Asta, A. Ingrosso, R. Zecchina, The patient-zero problem with noisy observations, Journal of Statistical Mechanics: Theory and Experiment P10016 (2014).

\bibitem[Lokhov 2014]{Lok14} A. Y. Lokhov, M. M\'{e}zard, H. Ohta, L. Zdeborov\'{a}, Inferring the origin of an epidemic with a dynamic message-passing algorithm, Physical Review E {\bf 90}, 012801 (2014).

\bibitem[Zanga 2014]{Zan14} W. Zanga, P. Zhang, C. Zhoub, L. Guob, Discovering multiple diffusion source nodes in social networks, Procedia Computer Science {\bf 29}, pp.443-452 (2014).

\bibitem[Buscema 2013]{Bus13} M. Buscema, E. Grossi, A. Bronstein, W. Lodwick, M. Asadi-Zeydabadi, R. Benzi, F. Newman, A new algorithm for identifying possible epidemic sources with application to the German escherichia coli outbreak, ISPRS International Journal of Geo-Information {\bf 2}, pp.155-200, (2013).

\bibitem[Thomas 2013]{Toh13} S. M. Thomas, C. Beierkuhnlein, Predicting ectotherm disease vector spread - benefits from multidisciplinary approaches and directions forward, Naturwissenschaften {\bf 100}, pp.395-405 (2013).

\bibitem[Maeno 2013]{Mae13} Y. Maeno, Transient fluctuation of the prosperity of firms in a network economy, Physica A {\bf 392}, pp.3351-3359 (2013).

\bibitem[Le Comber 2012]{LeC12} S. C. Le Comber, M. D. Stevenson, From Jack the Ripper to epidemiology and ecology, Trends in Ecology \& Evolution {\bf 27}, pp.307-308 (2012).

\bibitem[Le Comber 2011]{LeC11} S. C. Le Comber, D. K. Rossmo, A. N. Hassan, D. O Fuller, J. C. Beier, Geographic profiling as a novel spatial tool for targeting infectious disease control, International Journal of Health Geographics {\bf 10}, 35 (2011).

\bibitem[Cesar Henrique 2011]{CesC11} C. Cesar Henrique, L. da Fontoura Costa, Identifying the starting point of a spreading process in complex networks, Physical Review E {\bf 84}, 056105 (2011).

\bibitem[Maeno 2011]{Mae11} Y. Maeno, Discovery of a missing disease spreader, Physica A {\bf 390}, pp.3412-3426 (2011).

\bibitem[Maeno 2010]{Mae10} Y. Maeno, Discovering network behind infectious disease outbreak, Physica A {\bf 389}, pp.4755-4768 (2010).

\bibitem[Barrat 2004]{Bar04} A. Barrat, M. Barth\'{e}lemy, R. Pastor-Satorras, and A. Vespignani, The architecture of complex weighted networks, Proceedings of the National Academy of Sciences USA {\bf 101}, pp.3747-3752 (2004).

\bibitem[Hufnagel 2004]{Huf04} L. Hufnagel, D. Brockmann, and T. Geisel: Forcast and control of epidemics in a globalized world, Proceedings of the National Academy of Sciences USA {\bf 101}, pp.15124-15129 (2004).

\bibitem[Lipsitch 2003]{Lip03} M. Lipsitch {\it et al.}, Transmission dynamics and control of severe acute respiratory syndrome, Science {\bf 300}, 1966-1970 (2003).

\bibitem[Riley 2003]{Ril03} S. Riley {\it et al.}: Transmission dynamics of the etiological agent of SARS in Hong Kong: Impact of public health interventions, Science {\bf 300}, 1961-1966 (2003).

\bibitem[Canter 2000]{Can00} D. Canter, T. Coffey, M. Huntley, C. Missen, Predicting serial killers' home base using a decision support system, Journal of Quantitative Criminology {\bf 16}, pp.457-478 (2000). 

\bibitem[Rossmo 1993]{Ros93} D. K. Rossmo, A methodological model, American Journal of Criminal Justice {\bf 17}, pp.1-21 (1993).

\end{thebibliography}
\end{document}